\documentclass[aps,prl,twocolumn]{revtex4-2}

\usepackage{amsmath,amssymb,amsfonts,bm}
\usepackage{graphicx}
\usepackage{physics}
\usepackage{mathtools}

\begin{document}

\title{Neural Polaron: Learning Quasiparticle Operators in Quantum Many-Body Systems}

\author{Shang-Shun Zhang}
\affiliation{Department of Physics and Astronomy, University of Tennessee, Knoxville, Tennessee 37996, USA}

\date{\today}

\begin{abstract}
Understanding dynamical properties of quantum many-body systems remains a central challenge because excitations generally require information beyond a ground-state wave function. Here we introduce a neural polaron ansatz that represents quasiparticle excitations by neural many-body operators acting on a correlated ground state. Instead of learning an independent excited-state wave function, the method parameterizes a local dressing operator through a compact neural head defined on the feature map of a pretrained ground-state network. This operator-based construction builds in translation symmetry, momentum resolution, and quasiparticle locality, while separating ground-state correlations from excitation-specific dressing. We benchmark the method on the square-lattice $J_1$--$J_2$ Heisenberg model, where it accurately reproduces magnon dispersions and spectral weights over a broad range of frustration. In particular, it captures nontrivial many-body features including the $(\pi,0)$ anomaly and its progressive softening with increasing $J_2/J_1$. These results establish neural operators as a physically transparent route for extending neural quantum states from ground-state properties to dynamical response.
\end{abstract}

\maketitle

\textit{Introduction.---}
Complexity is a defining feature of quantum many-body systems, rooted in the exponentially large Hilbert space, strong interactions, and nontrivial entanglement \cite{Anderson1972, Preskill2012, Eisert2010,schollwock2011density,orus2014practical,sachdev1999quantum,sachdev2023quantum,wen2004quantum,abanov2003quantum,abanov2000spin,sandvik2010computational,troyer2005computational,laurell2025witnessing,stavropoulos2024complex}. 
While it underlies emergent phenomena without classical analogues, it also limits quantitative understanding of strongly correlated materials. Neural-network-based approaches have recently emerged as a powerful framework for representing such complex quantum states~\cite{carleo2019machine,carrasquilla2020machine,you2018machine,szabo2020neural,sun2022entanglement}. Starting from restricted Boltzmann machines \cite{Carleo2017} and extending to deep architectures~\cite{choo2019study,viteritti2026approaching,geier2025self,lou2024neural,nutakki2025design,viteritti2023transformer,rende2026transformer}, neural quantum states provide flexible variational representations capable of encoding long-range correlations and nontrivial entanglement structures.

Most applications of neural quantum states have focused on ground-state properties. Dynamical properties---including excitation spectra and spectral weights---are equally central, because they directly connect theory to spectroscopic probes such as inelastic neutron scattering and light-scattering experiments~\cite{chatterji2006magnetic,christianson2008unconventional,tennant2019studies,damascelli2003angle,Ament2011,xu2025ramification,park2026superconductivity}. They are also substantially more challenging. Ground states benefit from a variational principle and often obey comparatively restrictive entanglement structures~\cite{Eisert2010}, whereas excited states and real-time dynamics span a broader manifold with complex interference effects~\cite{Eisert2015}. Existing neural-network approaches to excitations typically either optimize independent excited-state wave functions subject to symmetry or orthogonality constraints~\cite{choo2018symmetries}, or pursue real-time evolution~\cite{mendes2023highly,ning2026recurrent}. These strategies are general, but they leave the physical structure of the excitation largely implicit in the wave function itself.

In this Letter, we introduce a complementary formulation: rather than using a neural network to represent the excited-state wave function directly, we use it to represent the quasiparticle operator. The excited state is constructed as a local, momentum-projected operator acting on a correlated ground state. This shift from neural wave functions to neural operators is the central idea of the neural polaron ansatz. It makes the quasiparticle structure explicit: locality, translation covariance, and momentum quantum numbers are built into the architecture, while the nontrivial dressing cloud is learned by a local neural ``polaron head.'' The many-body complexity is therefore separated into two parts. Ground-state correlations are encoded once in a pretrained backbone, while the excitation-specific variational problem is restricted to the local dressing operator. In the calculations below, the excitation head involves a compact subset of the full variational architecture. Its complexity can be systematically tuned through the head geometry, including the spatial kernel size, network depth, and number of channels. More importantly, the variational search is confined to a physically motivated quasiparticle manifold rather than the full space of excited-state wave functions.

We demonstrate this approach on the square-lattice $J_1$--$J_2$ Heisenberg model, accurately reproducing magnon dispersions and spectral weights across $0 < J_2/J_1 < 0.4$. The method captures subtle many-body features including the roton-like minimum near $\mathbf{q}=(\pi,0)$ and its progressive softening with increasing frustration~\cite{dalla2015fractional,powalski2018mutually,shao2017nearly,ferrari2018spectral,sandvik2001high,powalski2015roton,singh1995spin}. The optimized head also provides an physically interpretable probe of quasiparticle dressing: momenta and parameters that require a more expressive head correspond to excitations with stronger deviations from a bare spin flip. More broadly, the neural polaron ansatz provides an operator-based route for extending neural quantum states beyond ground-state calculations to momentum-resolved dynamical response.

\begin{figure}
    \centering
    \includegraphics[width=\linewidth]{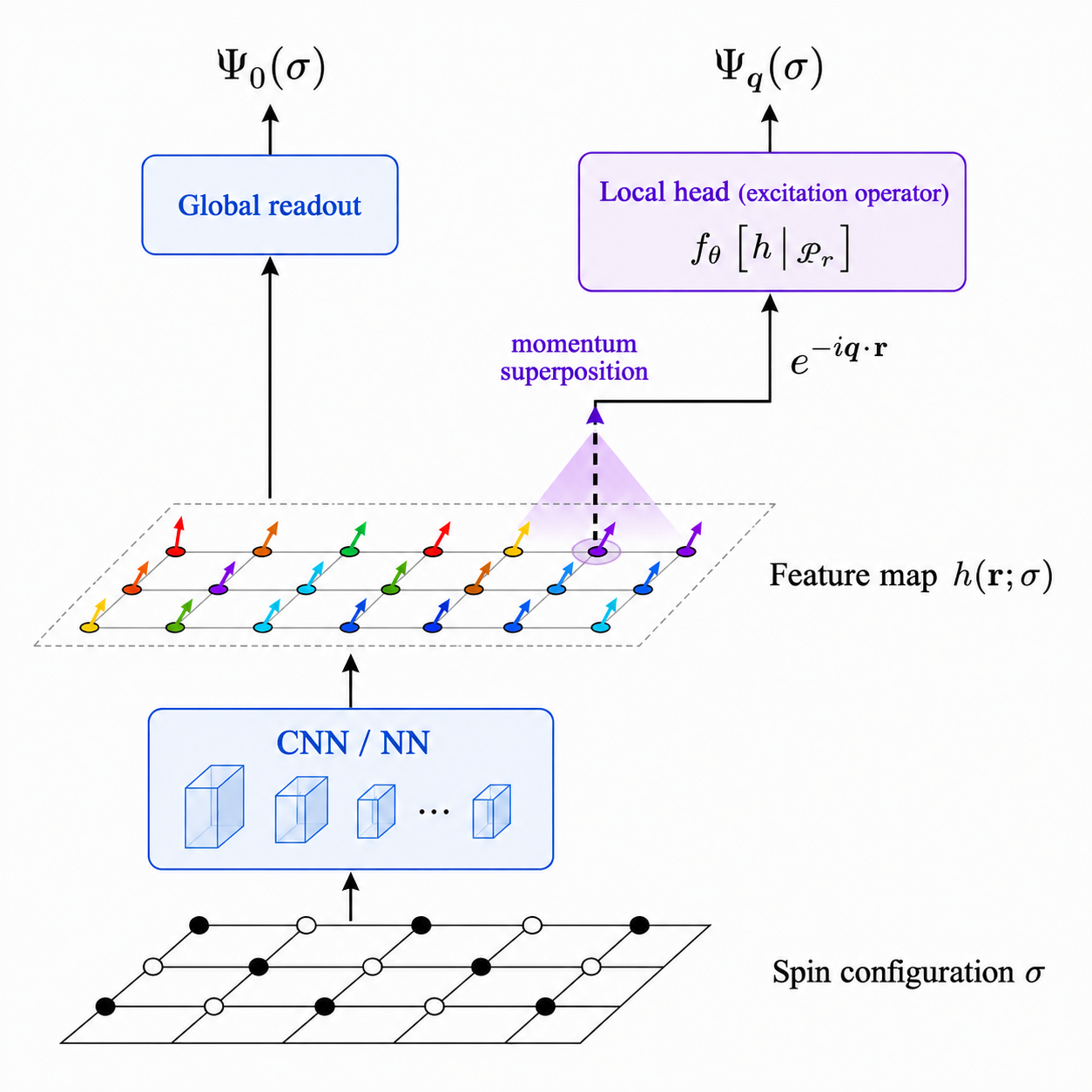}
    \caption{Schematic of the neural polaron ansatz. A ground-state neural network maps a spin configuration $\sigma$ to a spatial feature map $h(\mathbf{r};\sigma)$ that encodes many-body correlations. The same feature map is used to construct both the ground-state amplitude $\Psi_0(\sigma)$ via a global readout and quasiparticle excitations via a local neural head function $f_\theta$. The head acts on a finite spatial neighborhood of each site, defining a local excitation operator $O_{\mathbf{r}}\equiv f_\theta\!\left[h\big|_{\mathcal P_{\mathbf r}}\right]$, where  $\mathcal P_{\mathbf r}$ denotes a finite patch centered at $\mathbf r$. By applying the shared head across all lattice sites and superposing with a phase factor $e^{-i\mathbf{q}\cdot\mathbf{r}}$, the ansatz produces a momentum-resolved excited state $\Psi_{\mathbf{q}}$. This construction separates ground-state correlations from excitation dressing and restricts the variational space to a quasiparticle manifold.}
    \label{fig:illustration}
\end{figure}

\textit{Quasiparticle construction and variational ansatz.---}
A central organizing principle of quantum many-body systems is that low-energy excitations can often be understood as quasiparticles, i.e., collective modes built upon a correlated ground state~\cite{Landau_FL,tosatti2006quantum}. This suggests that one need not always represent excited states as independent wave functions in the full Hilbert space. Instead, one may construct them from the ground state through appropriately chosen many-body operators.

For a translationally invariant system, a momentum-resolved excitation can be written as
\begin{equation}
|\Psi_{\mathbf{q}}\rangle = \sum_{\mathbf{r}} e^{-i \mathbf{q}\cdot \mathbf{r}} \, O_{\mathbf{r}} \, |\Psi_0\rangle,
\label{eq:qp_ansatz}
\end{equation}
where $|\Psi_0\rangle$ is the ground state and $O_{\mathbf{r}}$ creates a local excitation centered at site $\mathbf{r}$. By construction, Eq.~(\ref{eq:qp_ansatz}) carries lattice momentum $\mathbf{q}$. The essential physical assumption is the \emph{locality} or \emph{quasi-locality} of $O_{\mathbf{r}}$: a quasiparticle is a localized disturbance dressed by correlations in its surrounding environment~\cite{devreese2009frohlich}.

This construction connects naturally to familiar limits. If $O_{\mathbf{r}}$ is a simple local operator, such as a spin flip in a magnetically ordered phase, Eq.~(\ref{eq:qp_ansatz}) reduces to the single-mode approximation~\cite{Feynman1954atomic,feynman1956energy,leggett2006quantum}. More elaborate linear combinations of nearby operators are consistent with spin-wave-type descriptions, where the excitation is weakly dressed by quantum fluctuations. In strongly correlated regimes, however, the dressing cloud can acquire a nontrivial spatial, amplitude, and phase structure that is difficult to capture by fixed analytic parametrizations. This motivates a flexible representation of the operator $O_{\mathbf{r}}$ itself.

\textit{Neural polaron ansatz.---}
Guided by Eq.~(\ref{eq:qp_ansatz}), our approach builds upon a neural representation of the ground state and learns the excitation operator as a local functional of the ground-state feature map (Fig.~\ref{fig:illustration}). We assume that the ground state $|\Psi_0\rangle$ is approximated by a neural-network ansatz that maps a spin configuration $\sigma$ to both a scalar amplitude $\Psi_0(\sigma)$ and a translation-covariant feature map $h(\mathbf r;\sigma)$. The feature map belongs to $\mathbb R^{N_c}$, where $N_c$ is the number of channels, and encodes the local and nonlocal correlations learned by the ground-state network.

The neural polaron wave function is written as
\begin{equation}
\Psi_{\mathbf{q}}(\sigma) =
\sum_{\mathbf r} e^{-i\mathbf q\cdot \mathbf r}
\, f_\theta\!\left[h\big|_{\mathcal P_{\mathbf r}}\right]
\, \Psi_0(\sigma),
\label{eq:neural_polaron}
\end{equation}
where $f_\theta$ is the neural polaron head and $h|_{\mathcal P_{\mathbf r}}$ denotes the restriction of the feature map to a finite patch centered at $\mathbf r$. The head therefore parameterizes the local operator coefficient associated with $O_{\mathbf r}$, while the correlated background is inherited from $\Psi_0$. Because the same head is shared over all lattice sites and then momentum-projected, the ansatz preserves translation symmetry and produces states with well-defined $\mathbf q$.

The key distinction from a generic excited-state neural quantum state is that the network does not have to rediscover the entire excited-state wave function. Instead, it learns a local dressing operator acting on a fixed correlated background. This operator-based inductive bias gives the ansatz a transparent physical interpretation. The patch size, depth, and number of head channels control the complexity of the dressing cloud. Even a minimal one-site head can capture nonlocal many-body correlations, because $h(\mathbf r;\sigma)$ itself inherits a finite receptive field from the ground-state backbone. Increasing the head expressivity then systematically enlarges the class of quasiparticle operators accessible within Eq.~(\ref{eq:neural_polaron}).

The parameters $\theta$ are optimized by minimizing the variational energy in the subspace defined by Eq.~(\ref{eq:neural_polaron}), using stochastic sampling over configurations $\sigma$. Additional symmetry or orthogonality constraints can be incorporated within the same framework~\cite{choo2018symmetries}. Since the ground-state backbone is fixed during excitation training, the optimization targets only the excitation-specific operator structure. This separation stabilizes the search and makes it possible to reuse the same correlated background for different momenta.

\begin{figure}
    \centering
    \includegraphics[width=\linewidth]{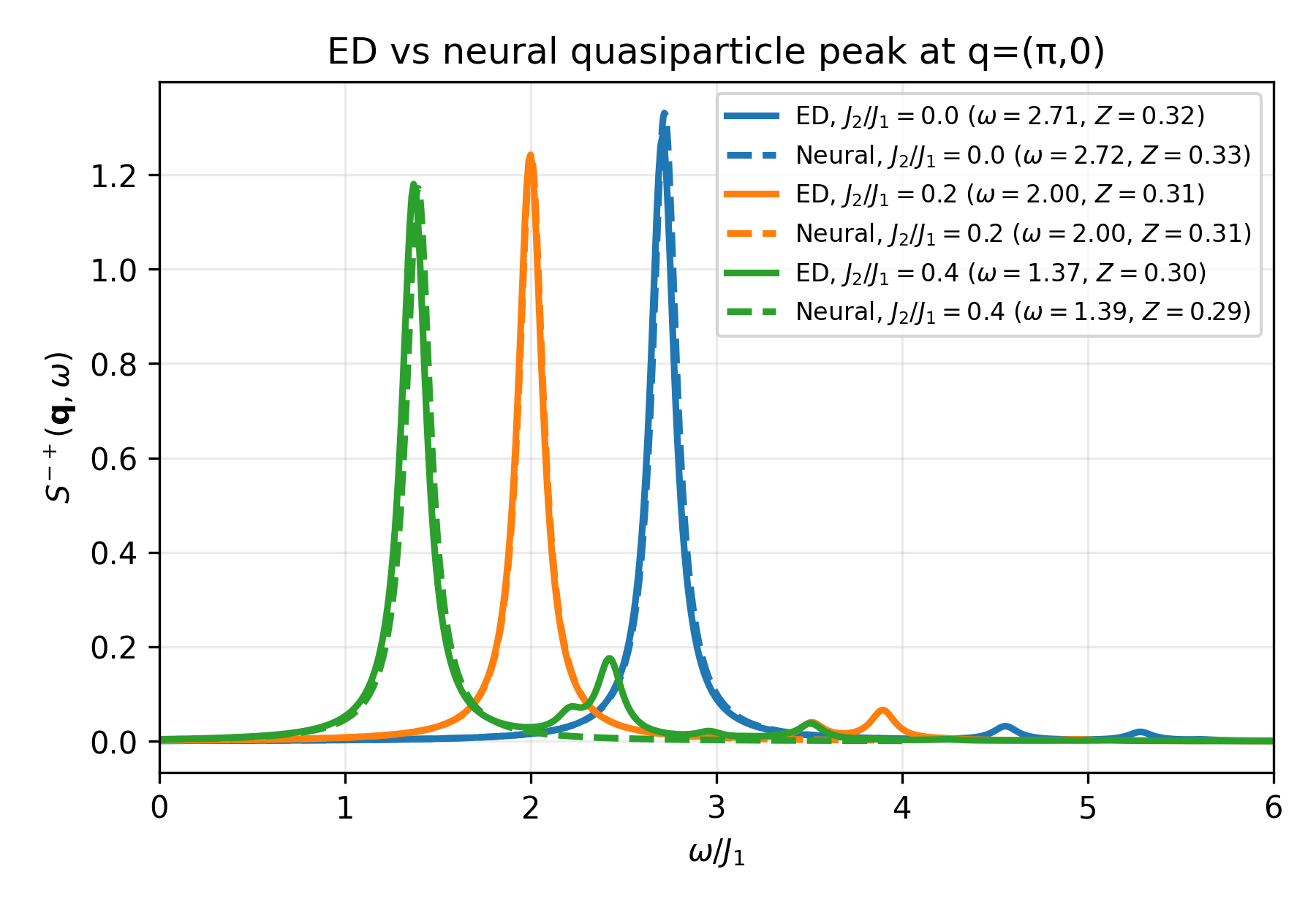}
    \caption{Exact diagonalization benchmark on a $4\times4$ lattice. Shown is the dynamical structure factor $S^{+-}(\mathbf{q},\omega)$ at fixed momentum $\mathbf{q}=(\pi,0)$ for $J_2/J_1=0$, $0.2$, and $0.4$. The neural polaron results (dashed lines) are compared with exact diagonalization (solid lines). Both the excitation energy and the spectral weight of the quasiparticle peak are accurately reproduced across the interaction range, demonstrating that the ansatz captures the position and residue of the magnon pole.}
    \label{fig:benchmark_ed}
\end{figure}

\textit{$J_1$--$J_2$ Heisenberg model.---}
We demonstrate the neural polaron ansatz on the square-lattice $J_1$--$J_2$ Heisenberg model,
\begin{equation}
H = J_1 \sum_{\langle ij \rangle} \mathbf{S}_i \cdot \mathbf{S}_j
  + J_2 \sum_{\langle\langle ij \rangle\rangle} \mathbf{S}_i \cdot \mathbf{S}_j,
\end{equation}
which provides a canonical platform for studying magnon excitations and their evolution from weak to intermediate coupling~\cite{shao2017nearly,sandvik2001high,powalski2018mutually,ferrari2018spectral,read1991large,chubukov1999behaviour,zhang2022schwinger}. We focus on the dynamical structure factor $S^{+-}(\mathbf{q},\omega)$, which directly probes momentum-resolved excitations (see Supplemental Material).

To illustrate the capability of the method, we consider the parameter regime $0<J_2/J_1<0.4$, where the ground state remains in the N\'eel phase while exhibiting progressively stronger correlations with increasing frustration. This regime therefore provides an ideal setting to test the evolution of quasiparticle dressing while retaining a well-defined quasiparticle description. The ground state is represented by a convolutional neural network (CNN)~\cite{liang2018solving,choo2019study,chen2022neural,roth2023high}, while the excitation operator is parameterized by a compact CNN head acting on the learned feature map. Details of the architectures are provided in the Supplemental Material.

\paragraph{Exact benchmark on $4\times4$.}
We first validate the method against exact diagonalization on a $4\times4$ lattice. Figure~2 shows the line cut of $S^{+-}(\mathbf{q},\omega)$ at fixed momentum $\mathbf{q}=(\pi,0)$ for representative values $J_2/J_1=0$, $0.2$, and $0.4$. Similar agreement is obtained at other representative momenta. The neural polaron ansatz accurately reproduces both the excitation energy and the spectral weight of the dominant quasiparticle peak across the entire interaction range. The remaining high-energy structures in the exact diagonalization spectra arise from multi-particle continuum excitations beyond the quasiparticle manifold considered here.

\begin{figure}
    \centering
    \includegraphics[width=\linewidth]{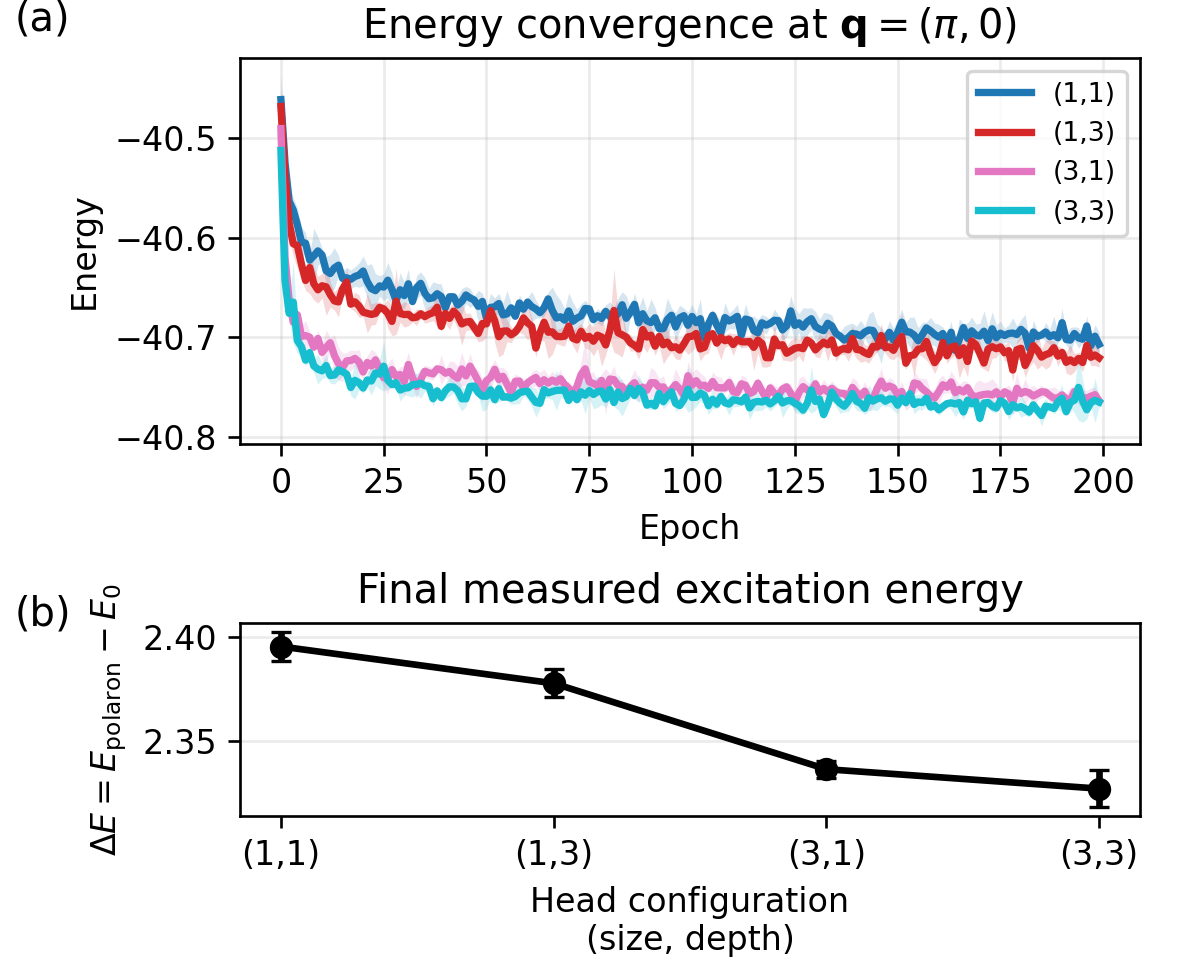}
    \caption{Expressibility of the neural polaron head function on an $8\times8$ lattice at $\mathbf{q}=(\pi,0)$ and $J_2/J_1=0$. 
(a) Energy versus training epoch for different head architectures characterized by kernel size and depth. Curves show seed-averaged results, with shaded regions indicating seed-to-seed variation. Increasing the spatial extent of the head systematically lowers the variational energy, indicating that a finite spatial dressing cloud is required for an accurate description of the $(\pi,0)$ quasiparticle.
(b) Estimated energy $E_{\mathrm{est}}$ as a function of head architecture $(\text{size},\text{depth})$. }
%The monotonic decrease and eventual saturation demonstrate that enlarging the receptive field is essential for capturing the extended quasiparticle dressing.}
    \label{fig:expressibility}
\end{figure}

\paragraph{Expressibility of the neural head.}
We next examine how the neural head captures quasiparticle dressing. We focus first on momentum $\mathbf{q}=(\pi,0)$, whose description is known to be particularly challenging and therefore provides a stringent test of the operator ansatz. Results for other momenta are presented in the Supplemental Material.

Figure~3(a) shows the energy convergence on an $8\times8$ lattice for different head architectures characterized by kernel size and depth. Enlarging the spatial support of the head produces the dominant energy improvement, whereas increasing depth at fixed kernel size gives a weaker gain. This trend is summarized in Fig.~3(b). The data show that a purely local head is insufficient to fully capture the excitation at $\mathbf q=(\pi,0)$, while a head with finite spatial extent substantially lowers the variational energy. Thus the head geometry provides a direct diagnostic of the spatial structure of the quasiparticle dressing cloud.

Among the architectures shown, a head with (kernel size, depth) $=(3,3)$ gives the best variational energy and provides the architecture used for the momentum-resolved spectra below. Preliminary calculations with a larger $(5,3)$ head converge more slowly and reach an energy comparable to, but not lower than, the $(3,3)$ head within the present optimization protocol. This suggests that the dominant quasiparticle dressing at $\mathbf q=(\pi,0)$ is already captured by a compact multi-site head, while larger heads may require more careful optimization. A systematic study of head geometry is left for future work.

A similar analysis at $\mathbf q=(\pi/2,\pi/2)$ (see Supplemental Material) reveals the same qualitative trend, although the energy gain from increasing the head expressivity is noticeably weaker than at $\mathbf q=(\pi,0)$. This momentum dependence indicates that the excitation near $(\pi,0)$ has a more pronounced dressing cloud and is farther from a simple bare spin flip.

To quantify this statement, we compare the optimized neural polaron state $|\Psi_{\mathbf q}\rangle$ with the seed excitation
\begin{equation}
|F_{\mathbf q}\rangle
=
S^-_{\mathbf q}|\Psi_0\rangle,
\qquad
S^-_{\mathbf q}
=
\sum_{\mathbf r}
e^{-i\mathbf q\cdot\mathbf r}
S^-_{\mathbf r},
\end{equation}
which corresponds to the simplest local spin-flip excitation. We define the normalized overlap
\begin{equation}
\tilde Z_{\mathbf q}
=
\frac{
|\langle \Psi_{\mathbf q}|F_{\mathbf q}\rangle|^2
}{
\langle \Psi_{\mathbf q}|\Psi_{\mathbf q}\rangle
\langle F_{\mathbf q}|F_{\mathbf q}\rangle
}.
\end{equation}
For $J_2=0$, we find $\tilde Z_{(\pi/2,\pi/2)}\simeq0.85$, compared with $\tilde Z_{(\pi,0)}\simeq0.72$. The reduced overlap at $(\pi,0)$ confirms a stronger departure from the bare spin-flip excitation, consistent with the larger gain obtained from a more expressive head.

\paragraph{Momentum-resolved dynamics.}
We now turn to the full dynamical structure factor on the $8\times8$ lattice. Figure~4 shows $S^{+-}(\mathbf{q},\omega)$ along the high-symmetry path $(\pi/2,\pi/2)\rightarrow(\pi,0)\rightarrow(\pi,\pi)\rightarrow(\pi/2,\pi/2)$ for representative values of $J_2/J_1$.

For $J_2=0$, the neural polaron results are in good agreement with quantum Monte Carlo (QMC) calculations~\cite{shao2017nearly}. In particular, the method correctly reproduces both the Goldstone mode near $\mathbf q=(\pi,\pi)$ and the characteristic downward renormalization near $\mathbf q=(\pi,0)$ associated with the $(\pi,0)$ anomaly. The remaining discrepancies are expected to originate primarily from finite-size effects, as evidenced by the evolution from the $L=4$ to $L=8$ results as well as the system-size dependence observed in the QMC study~\cite{shao2017nearly}. Beyond the dispersion, the momentum dependence of the spectral weight is also accurately reproduced, including the reduced quasiparticle weight near $\mathbf q=(\pi,0)$ relative to $(\pi/2,\pi/2)$, consistent with previous numerical and experimental studies~\cite{powalski2018mutually,dalla2015fractional}.

\begin{figure}
    \centering
    \includegraphics[width=\linewidth]{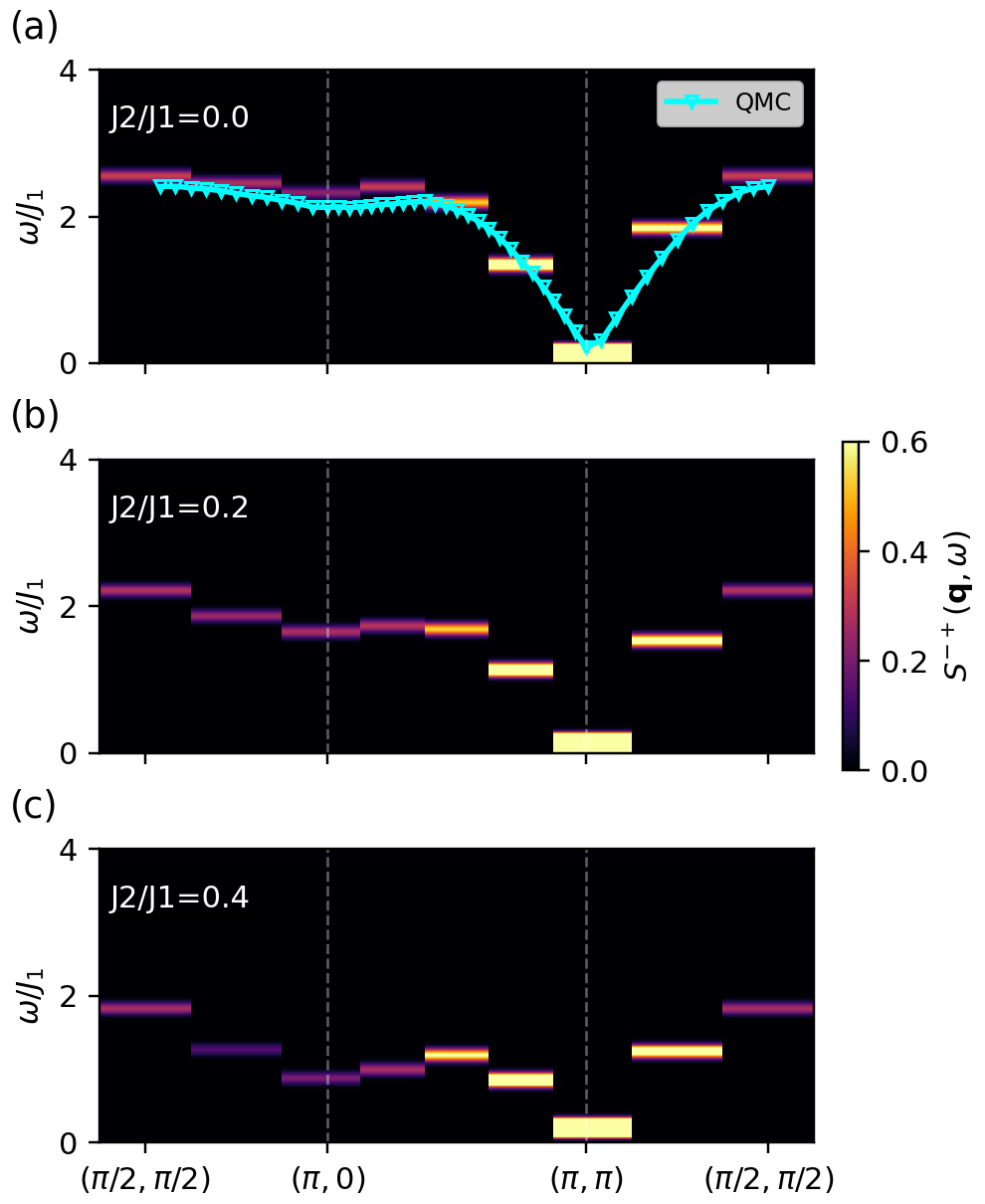}
    \caption{Momentum-resolved dynamical structure factor $S^{+-}(\mathbf{q},\omega)$ on an $8\times8$ lattice along the high-symmetry path $(\pi/2,\pi/2)\rightarrow(\pi,0)\rightarrow(\pi,\pi)\rightarrow(\pi/2,\pi/2)$ for representative values of $J_2/J_1$. Color maps show the neural polaron results. For (a) $J_2/J_1=0$, quantum Monte Carlo (QMC) data~\cite{shao2017nearly} is overlaid for comparison. The ansatz accurately reproduces the magnon dispersion and spectral weight, including the roton-like minimum near $\mathbf{q}=(\pi,0)$. As $J_2/J_1$ increases (see panel (b), (c)), the evolution of dispersion and intensity reflects the enhanced frustration and modification of quasiparticle dressing.}
    \label{fig:neural}
\end{figure}

As $J_2/J_1$ increases, both the dispersion and spectral weight evolve smoothly, reflecting enhanced frustration and increasingly nontrivial quasiparticle dressing. In particular, the excitation near $\mathbf{q}=(\pi,0)$ undergoes a pronounced softening accompanied by a substantial reduction of the quasiparticle overlap introduced above, from $\tilde{Z}_{(\pi,0)}\simeq 0.72$ at $J_2/J_1=0$ to $\tilde{Z}_{(\pi,0)}\simeq 0.67$ and $0.51$ at $J_2/J_1=0.2$ and $0.4$, respectively. The overall evolution of the spectra is consistent with previous studies based on Gutzwiller-projected variational wave functions~\cite{ferrari2018spectral}. These results indicate that the low-energy excitation remains continuously connected to a dressed quasiparticle throughout the N\'eel phase, while its internal structure becomes progressively more renormalized as frustration grows.

The pronounced softening and loss of quasiparticle weight near $(\pi,0)$ are consistent with the proximity of the N\'eel phase to the nonmagnetic regime of the square-lattice $J_1$--$J_2$ model. Recent studies have highlighted the close connection between this nonmagnetic regime and deconfined quantum criticality~\cite{wang2018critical,yang2022quantum,maity2603unified}. Within this picture, the $(\pi,0)$ anomaly may be viewed as a precursor of the enhanced quantum fluctuations associated with the nearby nonmagnetic regime, which manifest themselves through increasingly extended quasiparticle dressing already within the N\'eel phase.

\textit{Discussion and outlook.---}
The neural polaron ansatz introduced here provides an operator-based formulation of quasiparticle excitations in strongly correlated systems. Its central distinction from conventional neural excited-state approaches is that the neural network represents the excitation operator rather than an independent many-body wave function. The correlated ground state is encoded once in the backbone, while the excitation-specific structure is compressed into a local dressing head. This separation gives the method a transparent physical interpretation and restricts the variational search to a quasiparticle manifold with built-in locality and momentum resolution. Beyond providing a variational ansatz, the neural-polaron construction offers a way to extract physically interpretable information about the excitation itself through the learned operator structure. In this sense, neural networks are used not only to approximate quantum states, but also to learn quasiparticle operators and their associated dressing clouds.

The approach is broadly applicable to systems where a dominant quasiparticle pole carries significant spectral weight. For fermionic systems, the same construction naturally describes a dressed particle or hole propagating on a correlated background, closely paralleling conventional polaron descriptions and providing a potential route to momentum-resolved probes such as photoemission. The same idea can also be extended to electron--phonon models and materials with more complex interactions~\cite{choo2019fermionic,pfau2020ab,hermann2020deep,geier2025self,luo2026pairing}. More expressive architectures, including attention mechanisms or graph neural networks, could further capture nonlocal or anisotropic dressing structures.

At the same time, the method inherits the assumptions of a quasiparticle description. In regimes where the spectral weight is broadly distributed over continua, where quasiparticles are strongly damped, or where excitations fractionalize, a single local operator may no longer be sufficient. Extending the neural-polaron framework to multi-operator manifolds, multi-particle continua, or time-dependent variational subspaces is therefore an important direction for future work.

In summary, the neural polaron ansatz establishes a bridge between physically motivated quasiparticle constructions and modern neural-network representations. Beyond the present benchmark, it provides a complementary variational framework for studying dynamical properties in regimes where conventional numerical methods face distinct challenges. More broadly, the operator-based perspective introduced here opens a pathway toward physically interpretable neural descriptions of excitations in quantum many-body systems.

\textit{Acknowledgments --} 
The authors thank Hui Shao for sharing the QMC data used in the comparison and Anders W. Sandvik for valuable comments that helped improve the manuscript. This work used the Bridges-2 system at the Pittsburgh Supercomputing Center (PSC) through allocation PHY260014P from the Advanced Cyberinfrastructure Coordination Ecosystem: Services \& Support (ACCESS) program, which is supported by U.S. National Science Foundation grants \#2138259, \#2138286, \#2138307, \#2137603, and \#2138296. This work was supported by the U.S. Department of Energy, Office of Science, Basic Energy Sciences, Materials Sciences, and Engineering Division under award DE-SC-0018660.

\bibliographystyle{apsrev4-2} 
\bibliography{ref.bib}

%========================================================================================================================
% Supplemental Material
%========================================================================================================================

\clearpage
\onecolumngrid
\begin{center}
    \textbf{\large Supplemental Material for\\
    ``Neural Polaron: Learning Quasiparticle Operators in Quantum Many-Body Systems''}
    \\[10pt]
    Shang-Shun Zhang \\
    \textit{Department of Physics and Astronomy, University of Tennessee, Knoxville, Tennessee 37996, USA}
\end{center}

\setcounter{equation}{0}
\setcounter{figure}{0}
\setcounter{table}{0}
\setcounter{section}{0}

\renewcommand{\theequation}{S\arabic{equation}}
\renewcommand{\thefigure}{S\arabic{figure}}
\renewcommand{\thetable}{S\arabic{table}}
\renewcommand{\thesection}{S\Roman{section}}

\onecolumngrid

%========================================================================================================================
\section{Neural-network wavefunction architecture}
%========================================================================================================================

In this section, we describe the neural-network architecture employed for the square-lattice $J_1$-$J_2$ Heisenberg model. We focus on the antiferromagnetic N\'eel phase in the parameter regime
\begin{equation}
    0<J_2/J_1<0.4 .
\end{equation}
The wave functions are represented in the $S^z$ basis with periodic boundary conditions imposed on the square lattice.

%------------------------------------------------------------------------------------------------------------------------
\subsection{Ground-state neural quantum state}
%------------------------------------------------------------------------------------------------------------------------

The ground-state wave function is represented as
\begin{equation}
    \Psi_0(\sigma)
    =
    s_{\rm M}(\sigma)
    \exp[\Phi_\theta(\sigma)],
\end{equation}
where $\sigma=\{S_i^z\}$ denotes a spin configuration in the computational basis with
\begin{equation}
    S_i^z=\pm\frac12 .
\end{equation}
The factor
\begin{equation}
    s_{\rm M}(\sigma)
    =
    (-1)^{N^\downarrow_A(\sigma)}
\end{equation}
denotes the Marshall sign structure~\cite{marshall1955antiferromagnetism}, where
$N^\downarrow_A$ is the number of down spins on one sublattice of the bipartite square lattice.
The Marshall sign remains an excellent approximation throughout the N\'eel regime considered here, where the sign structure is still dominated by antiferromagnetic correlations~\cite{richter1994violation,retzlaff1993analysis}.

The scalar function $\Phi_\theta(\sigma)$ is parameterized by a convolutional neural network (CNN). The input spin configuration is mapped through stacked convolutional layers,
\begin{equation}
    h^{(n+1)}
    =
    f\!\left(
        W^{(n)} * h^{(n)}
        +
        b^{(n)}
    \right),
\end{equation}
where $*$ denotes convolution and $f(x)=\tanh(x)$ is the activation function. Periodic padding is imposed in each convolutional layer to preserve translation covariance. Residual connections are employed between convolutional blocks. The final output $\Phi_\theta(\sigma)$ is obtained through global pooling followed by a linear readout.

The CNN produces a translation-covariant feature map through the convolutional structure and periodic padding,
\begin{equation}
    h(\mathbf r;\sigma)
    \in
    \mathbb R^{N_c},
\end{equation}
where $N_c$ denotes the number of feature channels. Typical network parameters used throughout this work are: $N_c=16$, number of residual blocks $=6$, kernel size $=3$.

\begin{figure}
    \centering
    \includegraphics[width=0.6\linewidth]{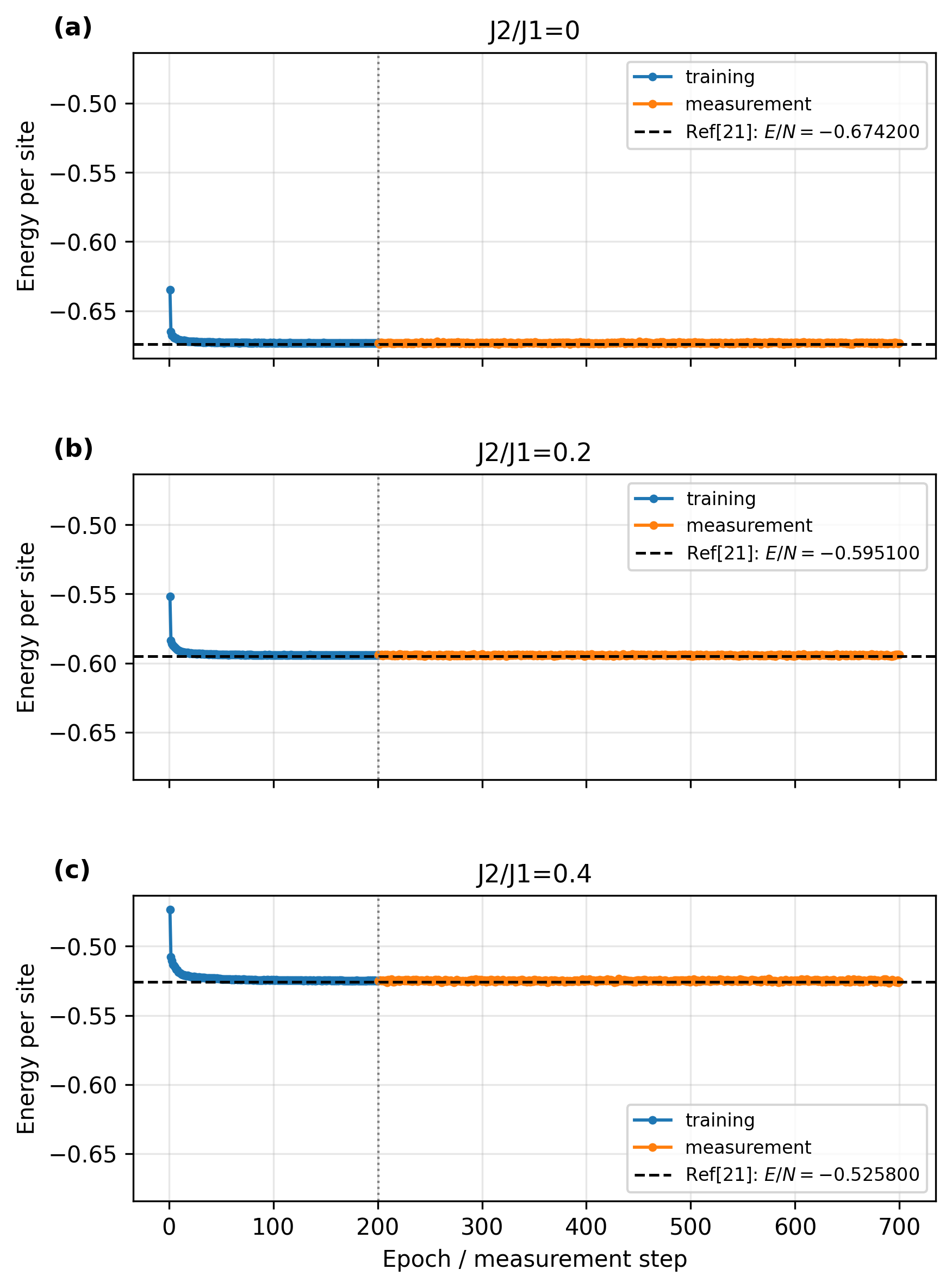}
    \caption{
    Ground-state energy per site as a function of optimization epoch for representative values of $J_2/J_1$ on the $8\times8$ square lattice.
    Blue and orange curves denote the training and measurement stages of the variational Monte Carlo optimization, respectively.
    Dashed horizontal lines indicate benchmark energies obtained from Ref.~\cite{choo2019study}.
    The neural-network ground-state ansatz converges rapidly and achieves excellent agreement with established numerical results throughout the antiferromagnetic N\'eel regime.
    }
    \label{fig:egs_convergence}
\end{figure}

In Fig.~\ref{fig:egs_convergence}, we show the training and measurement stages of the ground-state energy for representative values of $J_2/J_1$. In all cases, the variational ground-state energies are in excellent agreement with benchmark results from Ref.~\cite{choo2019study}, which employs a closely related CNN architecture differing mainly in implementation details such as the activation function, as well as finite-size DMRG calculations~\cite{gong2014plaquette} and Gutzwiller-projected variational Monte Carlo results~\cite{ferrari2018spectral}.

%------------------------------------------------------------------------------------------------------------------------
\subsection{Neural polaron head}
%------------------------------------------------------------------------------------------------------------------------

The ground state is optimized in the $S^z_{\rm tot}=0$ sector. The present work focuses on magnon excitations generated by transverse spin operators. In particular, we consider the excitation created by $S^-_{\mathbf q}$, which changes the total spin quantum number by
\begin{equation}
    \Delta S^z=-1 .
\end{equation}
Accordingly, the neural-polaron excitation wave functions are constructed and optimized in the $S^z_{\rm tot}=-1$ sector.

The excitation wave function is written as
\begin{equation}
\Psi_{\mathbf q}(\sigma)
=
\frac1{\sqrt N}
\sum_{\bm r}
e^{-i\mathbf q\cdot\mathbf r}
a(\mathbf r;\sigma^{+ \mathbf r})
\Psi_0(\sigma^{+\mathbf r}),
\label{eq:psiq_supp}
\end{equation}
where $\sigma^{+\mathbf r}$ denotes the parent configuration obtained by flipping the spin at site $r$ from down to up, satisfying
\begin{equation}
    S_{\mathbf r}^+\ket{\sigma^{+\mathbf r}}
    =
    \ket{\sigma}.
\end{equation}
The neural-polaron head acts on the feature map generated from the parent configuration $\sigma^{+\mathbf r}$.

The local excitation coefficient is parameterized as
\begin{equation}
    a(\mathbf r;\sigma)
    =
    \exp\!\left[
        A_\eta(\mathbf r;\sigma)
        +
        i\pi\tanh(
            \tilde\theta_\eta(\mathbf r;\sigma)
        )
    \right],
\end{equation}
where $A_\eta$ and $\tilde\theta_\eta$ are generated by a neural-polaron head network acting on the feature map $h(\mathbf r;\sigma)$ produced by the ground-state CNN. The head therefore outputs the logarithmic amplitude and phase of the local excitation coefficient.

\begin{figure}
    \centering
    \includegraphics[width=0.5\linewidth]{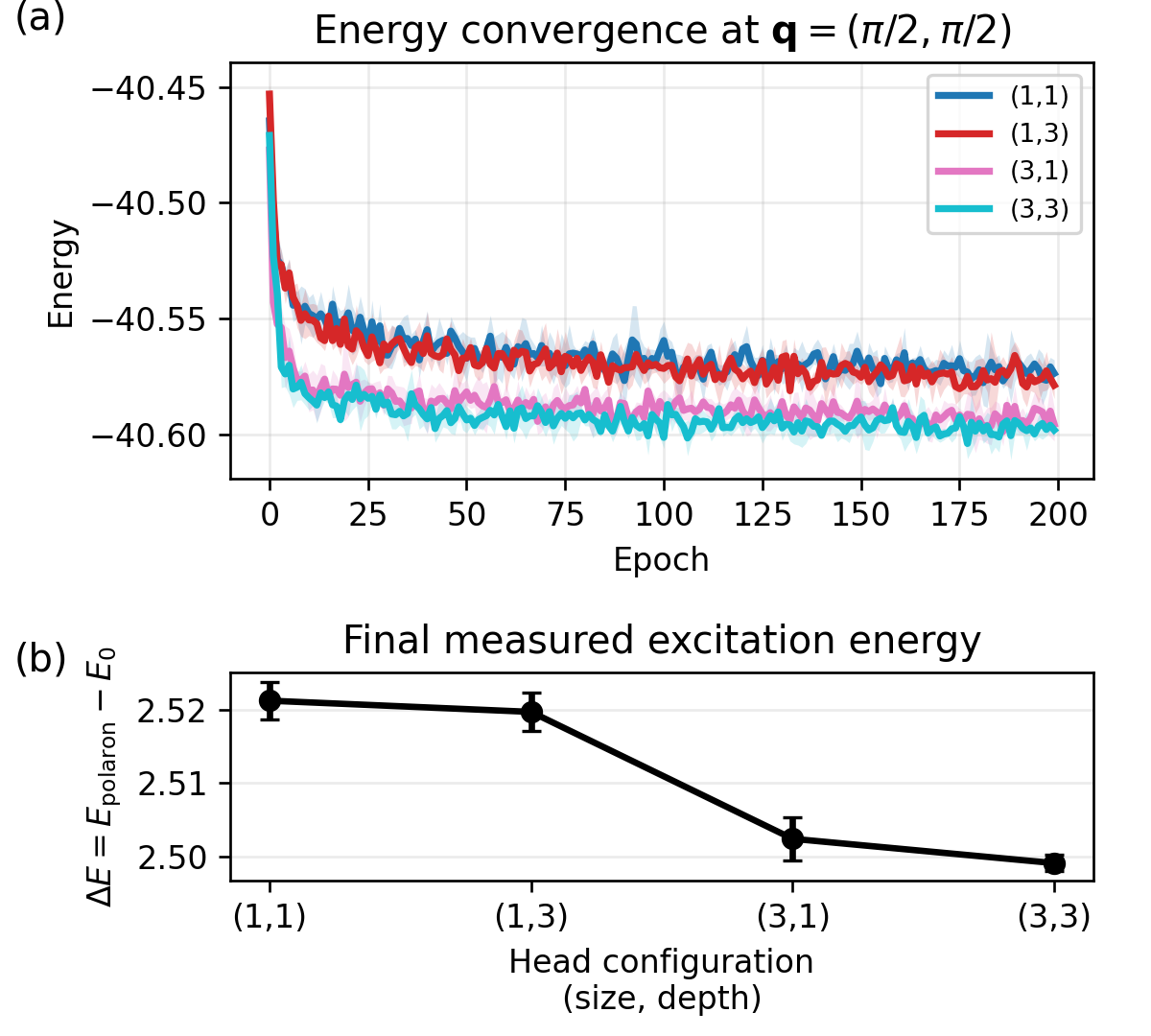}
    \caption{
    Expressibility of the neural-polaron head at momentum $\mathbf q=(\pi/2,\pi/2)$ on the $8\times8$ square lattice with $J_2/J_1=0$.
    (a) Energy convergence as a function of optimization epoch for different head architectures characterized by kernel size and depth: $(1,1)$, $(1,3)$, $(3,1)$, and $(3,3)$.
    Curves show seed-averaged energies, with shaded regions indicating seed-to-seed variations.
    (b) Final variational energy estimate $E_{\rm est}$ for each architecture.
    Increasing the spatial extent and depth of the head systematically improves the variational energy, demonstrating enhanced expressibility of the quasiparticle dressing structure.
    }
    \label{fig:convergence_head_config_J2_0p0_q22}
\end{figure}

The neural-polaron head is implemented as a compact CNN with periodic padding and $\tanh$ activations. The architecture is characterized by the kernel size $K$, network depth $D$, and number of hidden channels $N_h$. In the work, $N_h=64$, and typical head architectures include
\begin{equation}
    (K,D)
    =
    (1,1),
    (1,3),
    (3,1),
    (3,3).
\end{equation}
Increasing the kernel size and depth systematically enhances the expressibility of the neural-polaron head, as demonstrated in Fig.~\ref{fig:expressibility} of the main text for $\mathbf q=(\pi,0)$. A similar analysis for $\mathbf q=(\pi/2,\pi/2)$ is shown in Fig.~\ref{fig:convergence_head_config_J2_0p0_q22}. For the Goldstone mode at $\mathbf q=(\pi,\pi)$, a single-site head function $(1,1)$ already provides a near-optimal description. 

The effective receptive field scales approximately as
\begin{equation}
    \xi_{\rm head}
    \sim
    D(K-1)+1 .
\end{equation}
Even the minimal single-site head captures nonlocal many-body correlations, since the underlying feature map $h(\mathbf r;\sigma)$ already possesses a finite receptive field inherited from the ground-state CNN.

For reference, the number of excitation-specific parameters in the head scales approximately as
\begin{equation}
    N_{\rm head}
    \simeq
    K^2 N_c N_h
    +(D-1)K^2 N_h^2
    +N_{\rm out}N_h,
\end{equation}
up to bias terms and implementation-dependent readouts. Here $N_h$ is the number of hidden channels in the head and $N_{\rm out}=2$ for the logarithmic amplitude and phase outputs used here. For example, a $(K,D)=(3,3)$ head with $N_c=16$ and $N_h=16$ contains $O(10^4)$ excitation-specific parameters, whereas the more conservative $N_h=64$ choice used in the main benchmarks contains $O(10^5)$ parameters. In either case, the optimized parameters describe only the local quasiparticle dressing operator, while the ground-state correlations are inherited from the fixed backbone.

%========================================================================================================================
\section{Variational Monte Carlo optimization}
%========================================================================================================================

The neural-network wave functions are optimized using variational Monte Carlo (VMC). The variational energy is
\begin{equation}
    E
    =
    \frac{
        \langle\Psi|H|\Psi\rangle
    }{
        \langle\Psi|\Psi\rangle
    } .
\end{equation}
Monte Carlo sampling is performed using the probability distribution
\begin{equation}
    p(\sigma)
    =
    \frac{
        |\Psi(\sigma)|^2
    }{
        \sum_\sigma |\Psi(\sigma)|^2
    } .
\end{equation}
The energy can be expressed as
\begin{equation}
    E
    =
    \sum_\sigma
    p(\sigma)
    E_{\rm loc}(\sigma),
\end{equation}
where the local energy estimator is
\begin{equation}
    E_{\rm loc}(\sigma)
    =
    \sum_{\sigma'}
    H_{\sigma\sigma'}
    \frac{
        \Psi(\sigma')
    }{
        \Psi(\sigma)
    } .
\end{equation}

The variational parameters are optimized using the Adam optimizer with learning rate $\eta = 10^{-3}$. 
Typical calculations employ batches of $128$ Monte Carlo configurations for each parameter update, with $20$ optimization steps per epoch. The Monte Carlo sampling employs $200$ thermalization sweeps, followed by $200$ measurement sweeps for each optimization step. We additionally tested stochastic reconfiguration (SR) optimization~\cite{sorella2007weak} for representative cases and found results comparable to standard Adam optimization for the system sizes considered here.

To improve optimization stability, we use a soft tail-gating procedure that down-weights configurations with anomalously large local-energy deviations in the stochastic energy-gradient estimator. This procedure is controlled by robust batch energy scales and does not discard configurations outright. 

%========================================================================================================================
\section{Evaluation of dynamical spectral functions}
%========================================================================================================================

Using the optimized excitation wave functions, the dynamical spin structure factor is evaluated through the spectral representation
\begin{equation}
    S^{+-}(\mathbf q,\omega)
    =
    \sum_n
    Z_n(\mathbf q)
    \delta\!\left(\omega-\omega_n\right),
\end{equation}
where
\begin{equation}
    Z_n(\mathbf q)
    =
    \left|
    \left<
        n
        \middle|
        S^-_{\mathbf q}
        \middle|
        0
    \right>
    \right|^2
\end{equation}
denotes the spectral weight,
\begin{equation}
    S^-_{\mathbf q}
    =
    \frac{1}{\sqrt N}
    \sum_{\mathbf r}
    e^{-i\mathbf q\cdot\mathbf r}
    S^-_{\mathbf r}
\end{equation}
is the momentum-resolved spin operator, $S^-_{\mathbf r} = S^x_{\mathbf r} - i S^y_{\mathbf r}$ is the spin lowering operator under the $S_{\mathbf{r}}^z$ basis, and
\begin{equation}
    \omega_n
    =
    E_n-E_0
\end{equation}
is the excitation energy.

In the neural-polaron calculation, the quasiparticle residue is evaluated as
\begin{equation}
    Z_{\mathbf q}
    =
    \frac{
    \left|
    \left<
        \Psi_{\mathbf q}
        \middle|
        S^-_{\mathbf q}
        \middle|
        \Psi_0
    \right>
    \right|^2
    }{
    \langle\Psi_{\mathbf q}|\Psi_{\mathbf q}\rangle
    \langle\Psi_0|\Psi_0\rangle
    }  \equiv \tilde{Z}_{\mathbf{q}} S^{+-}(\mathbf{q}),
\label{eq:zq_supp}
\end{equation}
where 
\begin{equation}
    \tilde{Z}_{\mathbf{q}} = \frac{
    \left|
    \left<
        \Psi_{\mathbf q}
        \middle|
        F_\mathbf{q}
    \right>
    \right|^2
    }{
    \langle\Psi_{\mathbf q}|\Psi_{\mathbf q}\rangle
    \langle F_\mathbf{q}|F_\mathbf{q}\rangle
    }, \ \ \left| F_\mathbf{q}\right> = S^-_{\mathbf q}
        \left|
        \Psi_0 \right>
\end{equation}
is the normalized overlap introduced in the main text, and 
\begin{equation}
    S^{+-}(\mathbf{q}) = {\langle F_\mathbf{q}|F_\mathbf{q}\rangle \over \langle\Psi_0|\Psi_0\rangle}
\end{equation}
is the static spin structure factor of the ground state. For the exact diagonalization benchmark, the spectral weight is extracted from the overlap between the lowest excited state in the $(\mathbf q,S^z=-1)$ sector and the spin-flip state $S^-_{\mathbf q}|0\rangle$.

In practical calculations, the delta function is broadened using a Gaussian kernel
\begin{equation}
    \delta(\omega-\omega_n)
    \rightarrow
    \frac{
        1
    }{
        \sqrt{2\pi}\eta
    }
    \exp
    \left[
        -
        \frac{
            (\omega-\omega_n)^2
        }{
            2\eta^2
        }
    \right],
\end{equation}
where $\eta$ denotes the broadening width.

\end{document}